\documentclass[twocolumn,amsmath,amssymb,showpacs,pra]{revtex4-1} 
 
\usepackage{graphicx} 
\usepackage{bm}

\begin{document} 
 
\title{Time-reversal focusing of an expanding soliton gas in disordered replicas} 
 
\author{A. Fratalocchi$^{1,2}$}
 
\email{andrea.fratalocchi@kaust.edu.sa} 
\homepage{www.primalight.org}

\author{A. Armaroli$^3$ and S. Trillo$^3$}
 
\affiliation{ 
$^1$PRIMALIGHT, Faculty of Electrical Engineering; Applied Mathematics and Computational Science, King Abdullah University of Science and Technology (KAUST), Thuwal 23955-6900, Saudi Arabia\\
$^2$Dept. of Physics, Sapienza University of Rome, 
I-00185, Rome, Italy\\
$^3$Dipartimento di Ingegneria, Universit\`{a} di Ferrara, Via Saragat 1, 
44122 Ferrara, Italy
} 

\date{\today} 
 
\begin{abstract} 
We investigate the properties of time reversibility of a soliton gas, originating from a dispersive regularization of a shock wave, as it propagates in a strongly disordered environment. An original approach combining information measures and spin glass theory shows that time reversal focusing occurs for different replicas of the disorder in forward and backward propagation, provided the disorder varies on a length scale much shorter than the width of the soliton constituents. The analysis is performed by starting from a new class of reflectionless potentials, which describe the most general form of an expanding soliton gas of the defocusing nonlinear Schr\"odinger equation.  
\end{abstract} 
%\pacs{64.70.Pf, 75.10.Nr, 42.55.Zz} 
%\pacs{42.65.Jx, 42.65.Tg, 47.40.Nm, 52.35.Tc} %da prl exp shock 2009
 % from new scheme 2010
 \maketitle 

\section{Introduction}
Reversing the time dynamics of particles and waves has been always a central and fascinating theme in physics. 
While in classical thermodynamics irreversibility stems from the second principle, 
among several examples of macroscopic reversible dynamics 
time-reversal focusing (TRF) of classical waves has driven particular interest 
\cite{Derode95PRL.75.4206,deRosney02PRL.89.124301,Derode03PRL.90.014301,Fouque04PRL.92.094502,Lerosey04PRL.92.193904,
Garnier04PRL.93.154501,Papanicolaou04,Melkov06,Science07,Bartal09PRB.79.201103,reviewTR,Martin09PRL.100.044106,Barsi09}, generalizing
the concept of phase-conjugation \cite{Boyd} to non-monochromatic and transient excitation.
Generally speaking, a TR mirror is used to detect the far-field at given points and time-revert the wave components 
propagating from a source in an inhomogeneous (scattering) medium, 
thus producing a backward propagating replica of the wave-packet that focuses back at the source.
This phenomenon can occur regardless of the specific nature of the waves (i.e., for acoustic, electromagnetic, matter, elastic or seismic waves) provided dissipative effects remain negligible, and has applications ranging from super-resolution \cite{deRosney02PRL.89.124301,Science07,Bartal09PRB.79.201103,Barsi09} to enhancement of information transmission rate \cite{Science07},  
earthquake localization \cite{earthquake}, atom cooling \cite{Martin09PRL.100.044106}, image reconstruction \cite{Barsi09}, etc.\newline\indent
Though TR have been investigated mainly in the framework of linear propagation, it can be extended to {\em nonlinear} media \cite{Tanter01PRE.64.016602,Fouque04PRL.92.094502,Barsi09},
provided the governing model preserves time-inversion symmetry. 
However, in the nonlinear case a fundamental difficulty arises from the formation of classical shock waves, which break TR locally, 
leading to irreversible loss of energy across the shock (where dissipation is no longer negligible). 
Experiments \cite{Tanter01PRE.64.016602} and numerical calculations \cite{Fouque04PRL.92.094502} show indeed
that focusing can be recovered only when TR is performed before the shock distance. 
%\cite{NachbinSIAM} linear dispersive
At variance with such scenario, however, in many physical settings the regularizing mechanism of shocks
is constituted by {\em dispersion} (rather than viscous effects), which leads to dispersive shock waves (DSW) 
characterized by an expanding region filled with fast oscillations (modulated soliton trains) 
\cite{HauScience01,HoeferPRA06,WanNatPhys07,PhysRevLett.101.044101,PhysRevLett.102.083902,ArmaroliPRA09,NPWATM}. 
For integrable models with appropriate initial conditions such phenomenon can be described exactly in terms of a pure gas of soliton particles whose expansion becomes manifest beyond a point of wave-breaking determined by the collective behavior of solitons \cite{PhysRevLett.101.044101}.\\
\indent In this article we address the problem of TRF of a soliton gas generated beyond wave-breaking points 
in the presence of a disorder which breaks the integrability of the underlying model. 
Importantly, we explore the statistical stability of the TRF property, addressing the effect
of forward and backward nonlinear propagation occurring under different realizations of the disorder (with same statistical properties, i.e., two different replicas).
Since a general theory of multi-solitons in weakly-dispersive nonlinear media subject to random perturbations is not available, 
nor does the methods employed for a single soliton give useful insight for the gas ensemble  
\cite{PhysRep88,PhysRevLett.95.130403,PhysRevLett.100.144101,PhysRevLett.101.143903,PhysRevLett.103.210402}, 
we propose a novel statistical analysis by combining information theory \cite{Kullback51paper} 
with the replica approach of spin-glasses \cite{MPVBook,fischer93:_spin_glass}. 
In particular, we define a suitable \emph{Replica Entropy Overlap} (REO) probability function, 
which acts as a thermodynamic order parameter for the system. 
In the presence of disorder with a short correlation length, the REO predicts the emergence of a \emph{ferromagnetic} phase, associated with efficient TRF (beyond wave-breaking points), in contrast with correlation lengths larger than the soliton width, 
where the REO yields a pseudo-\emph{glassy} phase which entail spoiling of TRF properties.

The paper is organized as follows. In Section \ref{greys}, we introduce our model equation and analyze it in the limit of zero disorder, 
when the system is integrable. We show that a class of soliton gases can be characterized as reflectionless
potentials in terms of inverse scattering transform.  The shock evolution scenario in the integrable limit is discussed in Section \ref{shock0}. 
Section \ref{d00} is devoted to investigate the system dynamics in the presence of different types of disorder, with both slow and fast variation 
(different spatial degrees of correlation), assessing the properties of TRF and contrasting it with the behavior of linear waves.\\

\section{A grey-soliton gas}
\label{greys}
\indent We focus on a universal model of dispersive wave propagation in nonlinear media, namely the defocusing Nonlinear Scr\"odinger equation (NLS) \cite{HauScience01,HoeferPRA06,WanNatPhys07,PhysRevLett.101.044101}:
\begin{align}
\label{nls0}
 i \frac{\partial u}{\partial t}+\frac{\partial^2 u}{\partial x^2}-2(\lvert u \rvert^2-\rho^2)u= 2 R(x) u,
\end{align}
$R(x)$ being a general integrability-breaking perturbation.\\
We begin by considering Eq. (\ref{nls0}) in the integrable limit $R(x)=0$, with an input waveform 
%$u(x,0)=u_0(x)=-\rho(w+iv)[w\tanh(wx)+iv]$, being $w^2+v^2=1$. 
\begin{align} \label{input}
u(x,0)=u_0(x)=-\rho(w+iv)[w\tanh(wx)+iv],
\end{align}
with $w^2+v^2=1$. For $\rho=1$, the potential $u_0$ is the well-known one-soliton solution of the NLS with \emph{grayness} parameter $w^2$  controlling the maximum depth ($x=0$) and width of the dark (gray) notch propagating over the constant background with velocity $2v$ \cite{HMITTOS}. 
When $\rho \gg 1$ Eq. (\ref{input}) contains many solitons, similarly to the limit $v=0$ \cite{PhysRevLett.101.044101}.
However, the parameter $v \neq 0$ makes the inverse scattering approach
much more challenging than the particular case $v=0$, where the spectral transform can be given in closed form \cite{PhysRevLett.101.044101}.
Nevertheless the essential information, namely the discrete spectrum (soliton content), can be extracted as follows.
Following the approach of Ref. \cite{HMITTOS} we start from the linear auxiliary (Zhakarov-Shabat) problem of the NLS, 
with spectral parameter $\lambda$ (eigenvalue):
\begin{align}
\label{lap}
  \bigg( \frac{\partial}{\partial x}+\begin{bmatrix}
    &i\frac{\lambda}{2} &-u^*_0(x) \\
    &u_0(x) &-i\frac{\lambda}{2}\end{bmatrix}
  \bigg)\begin{bmatrix}
    \psi_{11} &\psi_{12}\\
    \psi_{21} &\psi_{22}\end{bmatrix}=0,
\end{align}
and look for its (matrix) Jost solutions ${\bf \psi}={\bf \phi}^{\pm}(x)$, which are generated from the boundary values 
at $x=\pm \infty$ obtained from Eqs. (\ref{lap}) with $\partial/\partial x=0$ and $u_0=u_0(\pm \infty)$. 
Referring to the element $\phi_{11}^{-}(x)$ (the other elements can be found by means of symmetries of Eqs. (\ref{lap}) \cite{HMITTOS}), 
and applying the coordinate change $s=[1+\tanh (wx)]/2$ 
as well as the F-homotopic transformation $\phi_{11}^{-}=(s-1)^{-i\frac{k}{4w}}s^{-i\frac{k}{4w}}\varphi$ (we set $k=\sqrt{\lambda^2 - 4\rho^2}$), 
we reduce Eq. (\ref{lap}) to the Heun equation (with poles in $0,1,c=(w-iv)/2w, \infty$) for $\varphi$, 
defined by the following Riemann $\mathcal{P}$ symbol \cite{HDE}:
\begin{align}
\label{psy}
  \varphi=\mathcal{P}\begin{Bmatrix}
&0 &1 &c &\infty &\\
&0 &0 &0 &\alpha &s &q\\
&1-\gamma &1-\delta &2 &\beta &\\
\end{Bmatrix},
\end{align}
where $\gamma=\delta= 1-i k/2 w$, $\beta=-i k/2 w+\rho$, $\alpha=-i k/2 w-\rho$, $q$ being an \emph{accessory} parameter with cumbersome expression \cite{HDE}.
Therefore, we can express $\phi_{11}^-(x)$ from the full set of the 192 local solutions $Hl$ of Heun equation for $\varphi$. 
For $x\rightarrow -\infty$ ($s\rightarrow 0$), we obtain
\begin{align}
  \label{js}
\phi_{11}^{-}=(s-1)^{-i\frac{k}{4w}}s^{-i\frac{k}{4w}}Hl(c,q,\alpha,\beta,\gamma,\delta,s).
\end{align}
Then the scattering coefficients $a(\lambda)$ and $b(\lambda)$ of the reduced monodromy matrix 
(and hence the reflection coefficient $R(\lambda)=b/a$) should be calculated
from the analytical prolongation of $\phi_{11}^-$ to $x\rightarrow +\infty$ ($s\rightarrow 1$), according with the standard relation  
\cite{HMITTOS}:
\begin{equation}
\label{js0}
  \phi_{11}^{-}(s)=a(\lambda)\phi_{11}^{+}(s)+b(\lambda)\phi_{12}^{+}(s).
\end{equation}
However, the \emph{connection problem} associated with the Heun equation, i.e. the analytic prolongation of Heun functions, lies still unsolved \cite{HDE}. 
To overcome the problem and arrive at the discrete spectrum, we resort to expanding the Heun functions $Hl$ in Eq. (\ref{js})
in terms of Hypergeometric functions $_2F_1$ \cite{jcp_heun}:
 \begin{align}
  \label{js1}
\phi_{11}^{-}=(s-1)^{-i\frac{k}{4u}}s^{-i\frac{k}{4u}}\sum_ng_n[_2F_1(\alpha,\beta,\gamma-n,s)],
\end{align} 
with $g_n$ expansion coefficients. By using the known prolongation of each Hypergeometric function in $s\rightarrow 1$, 
we express the generic term in Eq. (\ref{js1}) as:
\begin{widetext}
\begin{equation}
  \label{js2}
_2F_1(\alpha,\beta,\gamma-n,s)= \frac{|\Gamma(\frac{i k}{2 w})|^2}{\Gamma(\rho)\Gamma(1-\rho)}F^{+}(s)+ 
F^{-}(s) \frac{\Gamma(\frac{k}{2i w})^2}{\Gamma\big(\frac{k}{2i w}-\rho\big)\Gamma\big(\frac{k}{2i w}+\rho\big)},
\end{equation}
\end{widetext}
$F^{\pm}(s)$ being two independent Hypergeometric functions near $s \rightarrow 1$. 
For a reflectionless potential containing just solitons both $a$ and $b$ go to zero for every $s$, 
a condition which requires the RHS of Eq. (\ref{js0}) becoming null independently from $s$. 
By exploiting Eq. (\ref{js2}) and well known properties of Gamma functions, this condition requires:
\begin{align}
\begin{cases}
\rho=N\\
\frac{k}{2iw}-\rho=-n,
\end{cases}
\end{align}
with $N$ and $n$ positive integers.
These conditions, along with the constraint that eigenvalues lie in the gap $|\lambda| \le 2\rho$ \cite{HMITTOS}, 
yield $2N-1$ discrete eigenvalues (solitons):
\begin{align}
\label{eigs1}
  &\lambda_{\pm n}=\pm 2\sqrt{\rho^2-(\rho-n)^2w^2}, &\lambda_0=2v\rho
\end{align}
with $n=1,...,N-1$ ($\rho \equiv N$). In summary, the input field $u_0(x)$ corresponds, for integer $\rho$, 
to a transparent potential describing a gas of $2\rho-1$ soliton particles whose velocities (eigenvalues) are expressed by Eq. (\ref{eigs1}).
In the limit $v=0$, Eq. (\ref{eigs1}) correctly reduces to the spectrum already discussed in Ref. \cite{PhysRevLett.101.044101}.
%--------------- Figure 1 --------------------
\begin{figure}
\hspace{-0.6cm}
\includegraphics[width=9cm]{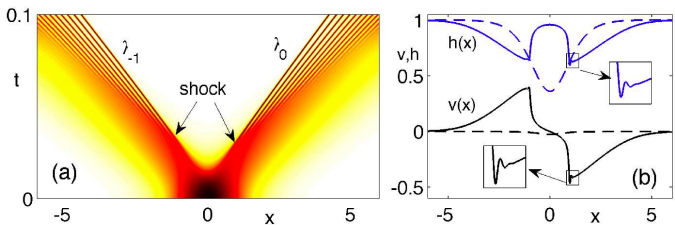}
\caption{
Color Online. (a) Level plot of $|u|^2$, showing the post-shock soliton expansion fans. 
The arrows indicate the two shock points as well as the slowest solitons (inner edges of the dispersive shock fans); 
(b) Snapshots of equivalent gas dynamic variables ${\rm v}$ and ${\rm h}$ vs. $x$ at $t=0$ (input, dashed lines) and $t=0.03$ (solid lines)
where the left shock develops, the right front already exhibiting the onset of oscillations, see insets. 
The results are from the numerical integration of the NLS equation in the integrable limit ($R=0$), 
with input $u_0(x)$, $v=0.6$ and $\rho=40$.}
\label{shock}
\end{figure}
%----------------------------------------------
\section{Shock scenario in the integrable regime}
\label{shock0}
In the strong multi-soliton regime ($\rho=N>>1$) of interest here, the soliton particles start to become manifest only beyond points of wave-breaking (gradient catastrophes),
where sharp gradients appear in the WKB variables:
\begin{equation}
\begin{cases}
h(x,t)=|u|^2/\rho^2,\\
v(x,t)=\frac{\partial_x arg(u)}{\rho},
\end{cases}
\end{equation}
% ${\rm h}(x,t)=|u|^2/\rho^2$ and ${\rm v}(x,t)=\rho^{-1} \partial_x arg(u)$
ruling the so-called dispersion-less limit of NLS \cite{HoeferPRA06,WanNatPhys07,PhysRevLett.101.044101,PhysRevLett.102.083902,ArmaroliPRA09}.
These variables obey, at leading order in $1/\rho$, 
hyperbolic equations of classical gas dynamic ($h$ and $v$ playing the role of density and velocity of the equivalent gas),
which predict the formation of two shocks (at slight different times) on opposite sides of the dark notch minimum \cite{ArmaroliPRA09}.
The integration of Eq. (\ref{nls0}) confirms this scenario (see Fig. \ref{shock}), 
showing a post-shock dynamics characterized by two asymmetric shock fans or DSWs,
which expand leftward and rightward, respectively. 
Inside these fans, solitons start to become manifest beyond the shock points, owing to regularization of the shock gradients due to dispersion.
They appear, starting from the two slowest (and hence darkest) ones with eigenvalues $\lambda_0=2N v$ for the rightward DSW, 
and $\lambda_{-1}=-2 \sqrt{N^2-(N-1)^2 w^2}$ (or $\lambda_{1}$ for $v<0$) for the leftward DSW (see arrows in Fig. \ref{shock}). 
These two asymmetric velocities define the inner edges of the two fans. 
The outer fan edges, conversely, are symmetric and correspond to shallow (quasi-linear) 
solitons with large velocities $\lambda_{\pm (N-1)}=\pm 2 \sqrt{N^2-w^2}$ (these are not yet visible in Fig. \ref{shock}). 
% new sentence
We point out that the knowledge of the edge velocities arises, in this case, from the explicit solution
of the direct scattering problem, while in non-integrable systems, in order to obtain the analogous information, 
one must perform lengthy calculations resorting to Whitham modulation theory.

\section{Time-reversal with disorder}
\label{d00}
Once the soliton gas propagates in the presence of disorder ($R \neq 0$), the expansion process of Fig. \ref{shock}
is perturbed with two main effects taking place: (i) the discrete spectrum changes (possibly including the creation of new eigenvalues), 
as can be proven by numerically calculating the set $\lambda_i(t)$, given $u(t,x)$ at any $t$ (see below);
(ii) a non-negligible amount of radiation is produced. 
As a result, the evolution could be hardly expected to be reversible when the wave-packet propagates backward in time over a different realization of the disorder.
In the following we investigate such issue, starting in Sec. \ref{n00} with numerical simulations aimed at showing the TRF behavior of a single DSW 
when it propagates over a sufficiently long time (i.e., such that the shock fans open up considerably).
After having highlighted the TRF features we propose in Sec. \ref{mb00} a statistical approach to study the phenomenon.
Lastly,  in Sec. \ref{h00}, we propose a simple approximate argument based on the Hamiltonian.  

\subsection{Numerical results}
\label{n00}
\begin{figure}
\includegraphics[width=9cm]{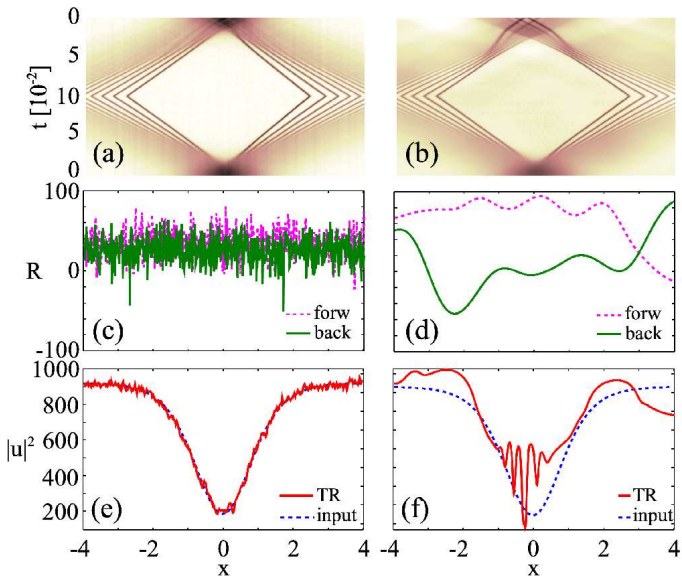}
\caption{
Color Online. Evolution of $\lvert u \rvert^2$ for two different $R(x)$ characterized by fast random (a)-(c)-(e) and slow pseudo-random (b)-(d)-(f) variations. 
The strength of disorder is equal ($\chi=0.5$) in the two cases.
(a-b) color level plot of the forward and backward propagation; (c-d) integrability breaking potential $R(x);$ (e-f) snapshots  of time-reversed (TR) output (solid lines)
compared with the input (dashed lines). Here the input is relative to $\rho=30$ and $w=0.9$ ($v \simeq 0.44$).
}
\label{prop}
\end{figure}

\begin{figure}
\includegraphics[width=9cm]{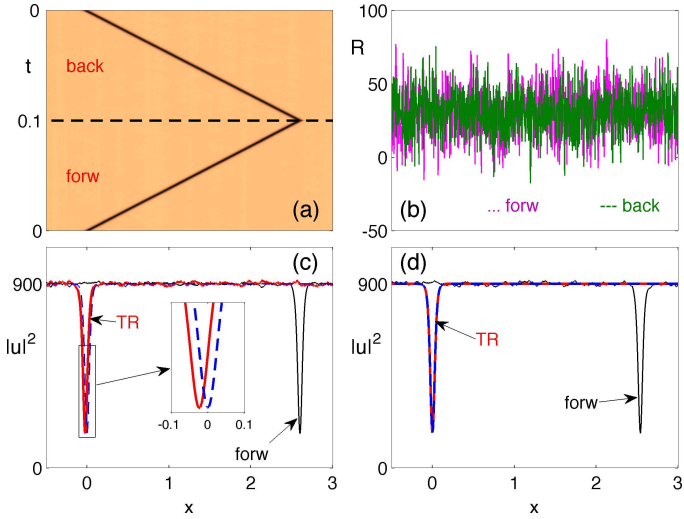}
\caption{
Color Online. Evolution of a 1-soliton with background $\rho=30$, and eigenvalue $\lambda_0=2v \rho$ ($w=0.9$, $v=\sqrt{1-w^2}$), 
in the presence of a fast random term $R(x)$ [same statistical properties as in Fig. \protect\ref{prop} (left column)]. 
(a) color level plot of the forward and backward propagation; 
(b) integrability breaking potential $R(x)$; (c) snapshots of TR (solid red line) and forward (thin solid black line) output,
compared with the input (dashed blue line). The inset shows the net displacement occurring upon TRF;
(d) reversed waveform for equal $R(x)$ in forward and backward propagation (ideal phase-conjugation): the TR output
is exactly superimposed to the input.
}
\label{sol}
\end{figure}

\begin{figure}
\includegraphics[width=8cm]{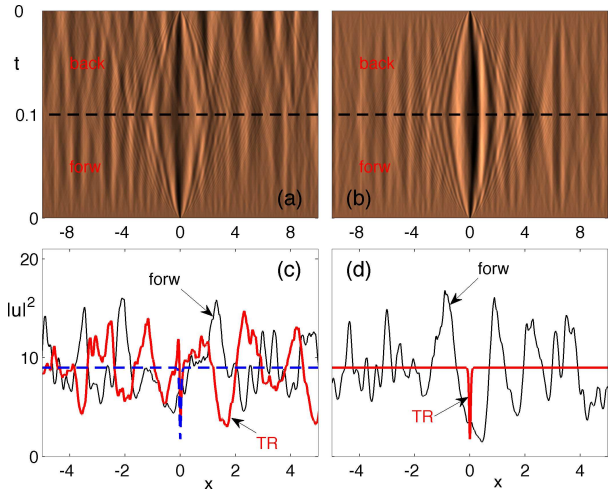}
\caption{
Color Online. Evolution of $\lvert u \rvert^2$ for a {\em linear} wave-packet (same waveform as in Fig. 3, 100 times weakened $|u|^2$) under $\delta$-correlated $R(x)$ 
[same statistical properties as in Fig. 2(left column) and Fig. 3]. (a-c) TRF when backward propagation takes place under a different realization of noise 
[$R(x)$ not shown, similar to those of Fig. \ref{sol}(b)]; (b-d) ideal phase conjugation for the same realization $R(x)$ in forward/backward propagation.
}
\label{lin}
\end{figure}

We have performed numerical simulations based on Eq. (1) with $R(x) \neq 0$ in forward and backward propagation. 
Our results show a qualitatively different behavior depending on the characteristic spatial scale (correlation) of the disorder $R(x)$.
A typical outcome is shown in Fig. \ref{prop}, where we contrast by means of numerical simulations two qualitatively different cases:\\
(i) the case of a normally distributed random variable $R$ with:
\begin{equation}
\begin{cases}
<R>=\rho,\\
\sigma_p=\rho\chi,
\end{cases}
\end{equation}
with $\chi$ being a variable coefficient,
% mean $<R>=\rho$ and standard deviation $\sigma_p=\rho\chi$ ($\chi$ being a variable coefficient)
yielding correlation $<R(x)R(x')>=\sigma_p^2\delta(x-x')$;\\
(ii) the case of a slow (long correlation length) pseudo-random term which we choose as:
\begin{equation}
R(x)=\sum_ng_n(x-nX_n)
\end{equation}
%$R(x)=\sum_ng_n(x-nX_n)$
with $g_n=a_n\exp[-(x/\omega_n)^2]$ being Gaussian functions with normally distributed amplitude $a_n$:
\begin{equation}
\begin{cases}
<a_n>=0,\\
<a_n^2>=\sigma_p^2,
\end{cases}
\end{equation}
%($<a_n>=0$, $<a_n^2>=\sigma_p^2$)
and displacement $X_n$, waist $\omega_n$ uniformly distributed in $[0.15(1-\chi,0.15(1+\chi))]$ and $[1-\chi,1+\chi]$, respectively 
(other pseudo-random $R(x)$ have been also tested and yield similar results). 
As shown, a remarkably good TRF is obtained, in spite of propagation over a different replica, only in the former case (delta-correlated noise), 
whereas a slowly-varying disorder leads to a poor reconstruction with solitons scattered at random locations [Fig. \ref{prop}(b,f)].\\
%--------------- figure 2 --------------------
%----------------------------------------------
The physical mechanism that leads to successful TRF for the case of the $\delta$-correlated noise, is the fact that soliton
constituents self-average the noise when the characteristic spatial scale of the latter is much shorter than their characteristic width.
Under these conditions the velocities (eigenvalues) of the soliton constituents do not change significantly in spite of the non-perturbative
character of the random term (amplitude variations of $R(x)$ are comparable with the soliton background amplitude),
thereby allowing to preserve the overall coherence of the whole soliton gas under time-reversal. 
This can be appreciated from Fig. \ref{sol} where we show the effect of fast noise [same statistical properties as in Fig. \ref{prop}(b,d,f)]
on a single dark soliton constituent, i.e. a narrow soliton sitting on the same background $u(\pm \infty)=\rho=30$ of the DSW modeled in Fig. \ref{prop}.
As an example we have chosen a soliton with eigenvalue $\lambda_0=2v \rho$, reporting in Fig. \ref{sol}(a,c) the evolution
occurring in the presence of two different realizations of the random term $R(x)$ [displayed in Fig. \ref{sol}(b)]. 
From Fig. \ref{sol}(c) it is clear that the net effect of the random term is to add small random fluctuations to the soliton background
and a small displacement in the return position. Such features, being due to the statistical difference of the fluctuations in forward and backward propagation,
disappear when backward propagation occurs under the same statistical realization of the noise, which leads to perfect reconstruction 
as shown for comparison in Fig. \ref{sol}(d). However, even in the presence of different statistical realizations, we observe that the net change of velocity of the soliton
(which has obviously statistical nature) remains extremely small, e.g. in Fig. \ref{sol}(b) we measure a net change $\Delta v/v < 0.4 \%$.
This allows for recombining all the solitons at the right locations thus reproducing the reversal of the entire gas or DSW shown in Fig. \ref{prop}(a).
It is important, however, to emphasize that it is only the combination of two ingredients, namely fast fluctuations and soliton features, 
that makes possible to obtain a rather good TRF. In general non-soliton wave-packets do not share this feature. 
For instance, in the linear regime, where the evolution is dominated by linear spreading (diffraction), TRF is not observed.
In order to show this, we report in Fig. \ref{lin} the outcome of a simulation of the evolution of a linear wave-packet. 
For comparison we inject the same waveform as in Fig. \ref{sol}, though with a hundred times lower $|u|^2$, 
and evolve the wave-packet in the same medium (i.e., same statistical properties of $R(x)$), over the same temporal scale.
As shown in Fig. \ref{lin}(a,c), time-reversal turns out to be quite poor,  with fluctuations which become comparable with the main dark notch injected in the medium.
This must be contrasted, once again, with perfect time-reversal reconstruction which is observed when backward propagation occurs under the same realization of noise,
as reported in Fig. \ref{lin}(b,d) for comparison. 
%-------------- figure 3: one-sol (new)
%----
%--------------- Figure 4: linear regime (new) --------------------
%----------------------------------------------

\begin{figure}
\includegraphics[width=9cm]{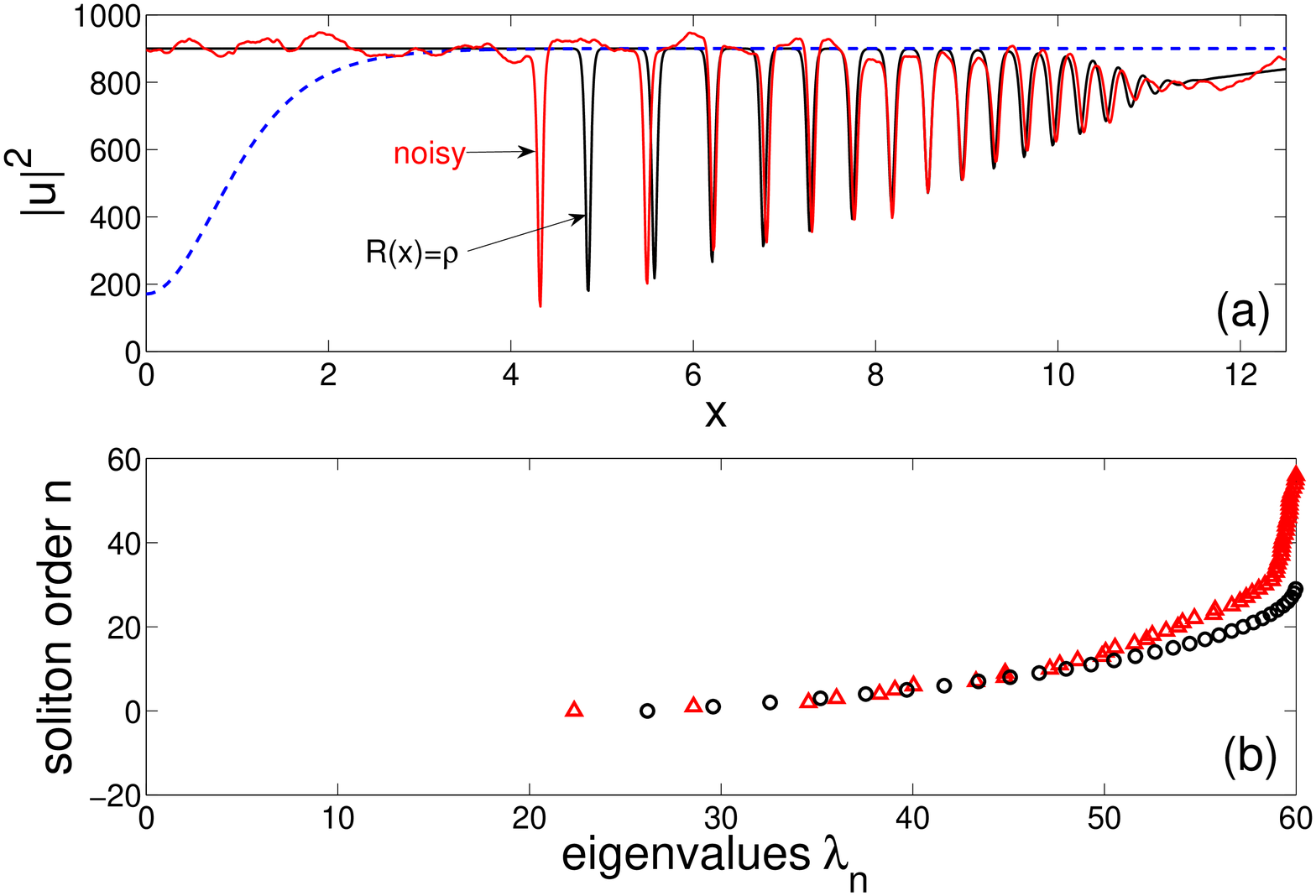}
\caption{
Color Online. Case of slow pseudo-random $R(x)$ [same statistical properties as in Fig. \protect\ref{prop}(b-d-f)]: 
(a) snapshots of the rightward expanding fan after forward propagation at $t=0.1$, comparing the noisy case (red solid curve)
with the integrable case (black solid curve, $R(x)=\rho$). The dashed blue curve stands for the input.
(b) relative soliton eigenvalues $\lambda_n$ (on the abscissa) vs. soliton number $n$ in the right half of the gap,
contrasting the integrable case (open circles) with the noisy one (triangles).
}
\label{eig}
\end{figure}

Conversely, when the characteristic length scale of $R(x)$ becomes comparable or larger than the width of soliton constituents of the DSW,
scattering of the solitons occurs according to the local features of the noisy potential $R(x)$. In this regime, significant statistical changes of the soliton velocities  
become important and lead to hamper the TFR feature. In order to deepen this point we have calculated the distribution of soliton velocities
by resorting to numerical calculation of the eigenvalues of the spectral problem [Eqs. (\ref{lap})], starting from the potential $u(x,t)$ obtained numerically.
To this end we have employed an algorithm based on Rayleigh-Ritz variational procedure \cite{PhysRevE.56.6142}, 
that we have adapted to deal with the defocusing case (self-adjoint scattering problem). 
In  Fig. \ref{eig}(b) we report the eigenvalues $\lambda_n$ (see triangles) vs. a progressive integer $n$ that identifies soliton components (soliton number) of the total field [shown in Fig. \ref{eig}(a)] emerging after forward propagation (at t=0.1), in the presence of $R(x)$ with long correlation length (as in Fig. \ref{prop}, right column). 
These data are contrasted with the situation such that random variations are switched off 
(open circles in Fig. \ref{eig}(b); for sake of comparison with the noisy case, we maintain $R(x)$ equal to the mean value $\rho=30$ and use the same algorithm,
though, obviously, the outcome is the same as in our analytical scattering approach in the integrable case, 
due to iso-spectrality and phase-rotation symmetry of the NLS equation).
By comparing the two sets of data reported in Fig. \ref{eig}(a,b), it is clear that significant random changes of velocities take place, especially for what concerns darker solitons,
which are expected to contribute more to the reconstruction of the input notch profile.
Also several new eigenvalues are created primarily at the edge of the spectral gap, where solitons turn out to be denser. 
Such alteration of the distribution of soliton velocities (and width and darkness as well, since they are related to velocity) 
is the clear signature of the fact that the noise is not simply averaged out in this case, thus spoiling TFR.
%-------------- figure 5 (new)
%----
\subsection{Replica theory: ferromagnetic vs glassy soliton phase}
\label{mb00}
\begin{figure}[h]
\includegraphics[width=9cm]{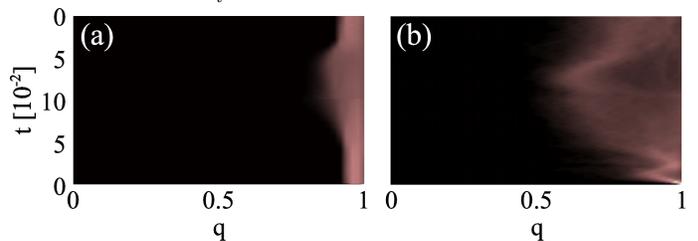}
\caption{
Color Online. Countour plot of entropy overlap probability $P(q)$ vs. time, obtained with disorder $R(x)$:
(a) uncorrelated (fast) random term with normal distribution;
(b) correlated pseudo-random term obtained through summation of Gaussians (see text);
The probability has been calculated by mediating over 20 samples ($\chi\in [0.01,0.5]$), each simulated with 20 different realizations of disorder. 
The input has been taken as a dense soliton gas with $\rho=30$ and $w=0.9$.
}
\label{reo}
\end{figure}

To proceed further and characterize the process of TRF statistically, we provide an approach based on the replica theory of spin-glasses. 
We consider two different observables of the system, $\{f^{(\alpha)}\}$, $\{f^{(\beta)}\}$, 
emerging from two different TRF experiments labeled $\alpha$ and $\beta$. 
Being interested in the TRF reproducibility of the dark notch, we pose $f=\lvert\rho-\lvert u \rvert\rvert$ (we remove the background $\rho$), 
normalizing so that $\int_{-\infty}^{+\infty} dx f(t,x)=1$. 
Then we define a suitable entropy distance $q_{\alpha,\beta}$ between $\{f^{(\alpha)}\}$, $\{f^{(\beta)}\}$:
\begin{equation}
\label{qab}
  [q_{\alpha,\beta}]^{-1} = 1+\frac{1}{2}[K(\alpha,\beta)+K(\beta,\alpha)],
\end{equation}
$K(\alpha,\beta)=\int dx f^{(\alpha)}\log\frac{f^{(\alpha)}}{f^{(\beta)}}$ being  the (non-negative) Kullbach information \cite{Kullback51paper}. 
By construction, $0\le q_{\alpha,\beta}\le 1$, with the extremes attained when the states $\alpha$ and $\beta$ turn out to be identical ($q_{\alpha,\beta} = 1$ or ferromagnetic phase), 
or viceversa uncorrelated ($q_{\alpha,\beta} = 0$, equivalent to a paramagnetic phase). Then in order to describe the statistics of all possible distances in the space of states,
we introduce the probability distribution function $P(q,t)$:
\begin{equation}
  \label{pq0}
  P(q,t)=<\sum_{\alpha,\beta\neq\alpha}\delta(q_{\alpha,\beta}-q)>,
\end{equation}
with $<...>$ denoting an ensemble average. It is the shape of this function $P(q,t)$, which we call \emph{replica entropy overlap}, 
that plays the role of a physical order parameter for the system. Figure \ref{reo} shows the behavior of $P(q,t)$, 
as calculated by averaging over $20$ samples with different degree of disorder (i.e. $\chi\in [0.01,0.5]$), 
each characterized by $20$ different specific realizations of $R(x)$. 
The number of samples and quenched replicas have been chosen in order to let $P(q,t)$ converge to a steady-state distribution. 
As shown, for a noise with short correlation length, the evolution exhibits a tendency to remain in a \emph{ferromagnetic} phase
characterized by a probability function peaked around $q \sim 1$. In this case the soliton wave-functions ``sample'' 
an average disorder during propagation and are therefore able to revert the evolution even in the presence of a different replica.
Viceversa, when the disorder $R$ has long correlation length, 
the temporal evolution of the system exhibits a clear tendency towards what is known as a \emph{glassy} phase \cite{fischer93:_spin_glass}, 
the latter being characterized by a space-of states filled continuosy with all the possible distances in $[0,1]$. 
In this case, irreversibility becomes manifest  by the fact that, upon backpropagation, $P(q,t)$ no longer return to be peaked at $q=1$, 
but continuosly fills the interval $\Delta q\in[0.5,1]$ (Fig. \ref{reo}).\newline\indent
%--------------- Figure 6 --------------------
%----------------------------------------------
\subsection{Hamiltonian perturbative analysis}
\label{h00}
In order to gain more physical insight on such behavior, we provide a simple though approximate argument. 
Starting from the Hamiltonian density $\mathcal{H}=\lvert u_x\rvert^2 + (\lvert u \rvert^2-\rho^2)^2 + 2R \lvert u \rvert^2$
of Eq. (\ref{nls0}), 
%\begin{align}
%\label{h0}
%  \mathcal{H}=\int\bigg[\lvert u_x\rvert^2 + \lvert u \rvert^4 + (2R-\rho^2)\lvert u \rvert^2 \bigg]dx,
%\end{align}
and assuming the randomness to be fast varying $R=R(x/\varepsilon)$ ($\varepsilon \ll 1$) and such that the form of each soliton is not appreciably modified, 
the Hamiltonian $H=\int \mathcal{H} dx$ becomes:
\begin{align}
\label{h1}
  H=\sum_i H_i+\int_{-\infty}^{+\infty}\lvert u \rvert^2R\bigg(\frac{x}{\epsilon}\bigg)dx,
\end{align}
where the sum represents a purely kinetic term given through the unperturbed Hamiltonian of the single soliton $H_i=\frac{1}{3} \left( 4\rho^2 - \lambda_i^2 \right)^{3/2}$. 
The evaluation of the last integral in Eq. (\ref{h1}) can be made in the spatial frequency domain as follows (hat denoting Fourier transform):
\begin{widetext}
\begin{align}
  \label{h2}
&\lim_{L\rightarrow\infty}\iint_{-L}^{+L}\lvert u \rvert^2\hat{R}\bigg(\frac{k}{\epsilon}\bigg)e^{i\frac{k}{\epsilon}x}dxdk= 
2\epsilon\rho^2\lim_{L\rightarrow\infty}\int\frac{\hat{R}}{k}
 \sin\frac{kL}{\epsilon}dk+O(\epsilon^2)=2\epsilon\rho^2\hat{R}(0)+O(\epsilon^2).
\end{align}
\end{widetext}
Equations (\ref{h1})-(\ref{h2}) entail that the dynamic is that of an ensemble of kinetic free particles,
perturbed, at leading order, by a term which depends only on the spatial mean of the disorder potential. 
The motion ruled by Eqs. (\ref{h1})-(\ref{h2}) is completely reversible and, more important, does not depend on the particular realization of $R$. This result relies on nature of the soliton wave-functions that, being more extended than the characteristic spatial scale of $R$, do not see the landscape of $R$ but just its average. This explains in simple terms why TRF occurs regardless of the specific configuration of disorder. In the case of slowly varying terms, conversely, Eq. (\ref{h2}) no longer holds and the soliton wave-functions evolve according with the valley and peak structure of $R$, thus spoiling statistical stability of TFR against different replicas.
\section{Conclusions}
In summary, we have presented an example of a reversible shock dynamics by considering the refocusing of an expanding soliton gas 
propagating over different replicas of a disordered channel. A successful recompression of the gas occurs for a fast disorder, 
thanks to the onset of a ferromagnetic phase in the order-parameter which characterize the process. 
The most obvious experimental test or our predictions can be made in nonlinear optics where DSW 
have been observed (though in settings better described by generalized NLS equations \cite{PhysRevLett.102.083902}) and $R(x)$ represents
a random refractive index distribution. However, the universality of the NLS model let us envisage also experimental tests in other fields,
for instance with matter waves in Bose-Einstein condensates.\\

\section{Acknowledgements}
This research has received funding by Award No. KUK-F1-024-21 (2009/2012) made by King Abdullah University of Science and Technology (KAUST).

%\bibliography{shockref} 

%Merlin.mbs v4.21 2009-07-09.
%

%------------------------------------------------------------
\end{document}